\documentclass[12pt,notitlepage,superscriptaddress,aps,prd,nofootinbib]{revtex4-1}
\usepackage[T1]{fontenc}

\usepackage[latin9]{inputenc}
\usepackage{amsmath,amssymb}
\usepackage{xcolor}
\usepackage{graphicx}
\usepackage{graphics}
\usepackage{hyperref}
\usepackage{setspace}
\begin{document}
\pagenumbering{gobble}
\clearpage
\begin{center}
\textit{Essay written for the Gravity Research Foundation 2022 Awards for Essays on Gravitation,\\
Honorable Mention}    
\end{center}

\title{The Cosmic Web Crystal:\\
Ising model for Large Scale Structures}
\author{Leonardo Giani\footnote{Corresponding author}}
\email{uqlgiani@uq.edu.au}
\affiliation{School of Mathematics and Physics, The University of Queensland, Brisbane, QLD 4072, Australia.}

\author{Tamara Maree Davis}
\email{tamarad@physics.uq.edu.au}
\affiliation{School of Mathematics and Physics, The University of Queensland, Brisbane, QLD 4072, Australia.}
\begin{abstract}If Dark Matter halos possess the gravitational equivalent of an intrinsic  magnetic spin, a formal analogy exists between the low redshift behaviour of the Cosmic Web in a flat FLRW background, and a crystal of spins submerged in a thermal reservoir with temperature $T \propto H(t)$.
We argue that, within the use of the Bianchi type IX geometry to describe the gravitational collapse of matter inhomogeneities, the spins are nothing but the heritage of its underlying $SU(2)$ symmetry. Therefore, just like electrons in quantum mechanics, these structures may have spin independently from their orbital angular momentum.
We explore the phenomenological implications on cosmological scales of a possible late time phase transition of the Cosmic Web towards (the gravitational equivalent of) a ferromagnetic state, described qualitatively using the Ising model in the mean field approximation.

\end{abstract}
\maketitle
\thispagestyle{empty}
\clearpage

The concordance model of Cosmology, the $\Lambda$CDM model, rests on a trinity of foundational pillars: that General Relativity (GR) describes the gravitational interaction, that two dark species beyond the standard model of particle physics contribute to most of the Universe's energy-momentum content, and that the Cosmological Principle (CP) is valid at sufficiently large scales.
State of art observational Cosmology, on the other hand, compels us to critically inspect these pillars and their realm of validity. For example, it seems difficult to achieve a quantum mechanical description of GR within the same framework of the other known fundamental interactions, which led many to believe that Einstein's Relativity is an effective description of some unknown underlying fundamental physics. Furthermore, the interpretation of the Dark sector within the tools of particle physics is facing a number of practical and theoretical issues, like the so far unsuccessful direct detection of Dark Matter (DM) in lab experiments, or the 106-years-old\footnote{According to Ref. \cite{CCproblemfather}, Nobel laureate Walther Nernst first aknowledged the existence of a divergent ``zero point vacuum energy'' in 1916.} Cosmological Constant (CC) problem.
\pagenumbering{arabic}
In this essay we are mostly concerned about the reliability of the third pillar. The CP states that the Universe as a whole is homogeneous and isotropic, and mathematically incorporates the somehow reassuring\footnote{As posited in the \textit{Galaxy Song} from the movie \href{https://www.imdb.com/title/tt0085959/}{\textit{Monty Python's The Meaning of Life}},  by Idle \& Du Prez.} Copernican principle, which claims that we are in no way ``special observers''. On the other hand, it is clear that the  CP is meaningful only in a statistical sense due to  the naked eye acknowledgeable existence of inhomogeneities.\footnote{Those, who do not trust their sight, may find convincing Descartes's argument:  \textit{Cogito ergo sum}.} In the standard model, these inhomogeneities are the result of the gravitational collapse of initially small perturbations of the FLRW Universe, fingerprinted in the temperature fluctuations of the Cosmic Microwave Background (CMB). The origin of these perturbations is still somehow mysterious, but it is commonly believed that they are the remnant of the quantum fluctuations of the Inflaton field (or fields) dominating the Universe prior to the Big Bang.

One of the most striking successes of the $\Lambda$CDM model is its ability to provide a consistent description of the evolution of these perturbations from the radiation dominated epoch, where they expand with the homogeneous background, through the matter dominated epoch, when sufficiently overdense regions undergo gravitational collapse. These, eventually, results in the formation of the large scale structures weaving the Cosmic Web. 
At linear order, our knowledge is excellent because the evolution equations and their solutions are analytical. Beyond the linear regime, however, our understanding relies on the results of N-body and hydrodynamic simulations. Even if less appealing from the theoretical point of view, these simulations are nevertheless in excellent agreement with the observed distribution of large scale structures and therefore support the $\Lambda$CDM paradigm.

In order to simplify the Einstein field equations, it is an old theorist's trick to invoke symmetry assumptions for the underlying spacetime. Whilst the CP is in itself one such assumption, we get some insight about the physics of the process of structure formation by postulating additional symmetries for the geometry of the inhomogeneities. Two well known models of gravitational collapse, valid beyond the linear regime, are the Spherical Collapse model and the Zeldovich solution. The former describes the collapse of a spherical overdense shell of matter, and the latter the collapse of a 1-dimensional perturbation. It was recently shown in \cite{Giani:2021gbs}, that these are nothing but particular cases of a more general geometric hypothesis, i.e.~that the collapse is of Bianchi IX (BIX) type. Indeed, demanding that the evolution of the inhomogeneity can be described in terms of a single time parameter (i.e.~the cosmic time of the FLRW background), we are left with the list of all the possible homogeneous but anisotropic geometries, i.e.~the Bianchi classification. Furthermore, we aim to model inhomogeneities which were initially expanding and eventually undergo gravitational collapse. This inevitably restricts our choice to spatially closed geometries, otherwise it would not be possible for them to detach from the Hubble flow and begin their contraction. Being the only member of the Bianchi classification satisfying this requisite, we are thus left with the BIX as the only suitable candidate.    

It is always possible to represent the line element of a Bianchi geometry in the following form \cite{Landau:1982dva}:
\begin{equation}\label{linelement}
    ds^2 = -dt^2 + a^2(t) \omega^1\otimes\omega^1+b^2(t) \omega^2\otimes\omega^2+
c^2(t) \omega^3\otimes\omega^3 \;,
\end{equation}
where $a,b,c$ are the scale factors,\footnote{In a FLRW we have $a=b=c$, that is, a single scale factor which parametrise the volume of the comoving portion of the Universe as $V \propto a^3$.} responsible for the homogeneous anisotropic expansion, and $\omega^i$ with $i = \{1,2,3\} $ are the Maurer-Cartan basis 1-forms \cite{CartanSurLS}.  Eq. \eqref{linelement} is written using a subclass of coordinates corresponding to a choice of non-commuting basis vectors (with the exception of the Kasner metric\footnote{Which describes the spatially flat Bianchi type I spacetime, for which the basis vectors commute.}). The resulting metric connection is not expressed in terms of the Christoffel symbols, but with a number of constant and non-vanishing unholonomity coefficients.
For BIX the Maurer-Cartan basis 1-forms are:
\begin{eqnarray*}
\omega^1 &=& -\sin{z}dx + \sin{x}\cos{z}dy\;, \\
\omega^2 &=& \cos{z}dx + \sin{x}\sin{z}dy\;,\\
\omega^3 &=& \cos{x}dy +dz\; ,
\end{eqnarray*}
where $x,y,z$ are the usual Cartesian coordinates. The corresponding dual vector basis $\xi_i$ satisfy the following algebra:
\begin{equation}
    \left[\xi_a, \xi_b\right]= \epsilon_{abc}\xi_c \;, 
\end{equation}
where $\epsilon_{abc}$ is the Levi-Civita symbol. Hence, the underlying symmetry of the BIX spacetime is the non-abelian Lie group $SU(2)$, i.e.~the spin group. Newborn theoretical physicists are usually taught, in their QFT courses, to take symmetries very seriously. Indeed, particles and interactions emerge as representations of their corresponding symmetry groups. Quoting the Nobel laureate Steven Weinberg: \textit{The universe is an enormous direct product of representations of symmetry groups.
}  In this essay we explore the possibility that large scale structures, which we assumed have  originated through BIX gravitational collapse, carry an associated intrinsic spin number. We stress that the latter is not to be interpreted as the classical, orbital angular momentum of a rotating object, but as an intrinsic ``quantum'' number analogue to the spin of a point-like electron,\footnote{More precisely, we are defining as ``intrinsic'' spin the one related to the irreducible representations of the $SU(2)$ algebra, which from a QFT perspective corresponds to the spin of the elementary particles. It has been shown that BIX can be used to describe the approach to the cosmological singularity \cite{Landau:1982dva}, and in this point-like and non-rotating limit the associated $SU(2)$ conserved charge cannot be ascribed to its orbital angular momentum.} which sums to the former to give the total angular momentum.  Our focus is on the  phenomenological implications on the Cosmic web which could arise from their  spin-spin interactions. Ultimately, the motivation for our investigation is more easily understood (again) with the words of Professor Weinberg: \textit{Nothing in physics seems so hopeful to me as the idea that it is possible for a theory to have a very high degree of symmetry  which is hidden from us in ordinary life.\footnote{\href{https://doi.org/10.2307/3823787}{https://doi.org/10.2307/3823787}}}

So far, we focused our presentation on the description of the matter inhomogeneities. Now, let us briefly turn our attention to the flat, isotropic and homogeneous background Universe. Undeniably, one of the most fascinating aspects of the gravitational interaction is its deep connection with the laws of thermodynamics. Standing on an Einstein elevator, free falling towards a black hole, and staring at the window (usually trying to remember where we parked the night before), we perceive empty space as a thermal bath with Davies-Unruh temperature $T_{DU}$ \cite{Davies_1975,PhysRevD.14.870}. Even if less spectacular, a similar phenomena occurs if we remove the black hole and consider instead free falling elevators in a spatially flat FLRW Universe with dust and a cosmological constant. Indeed, it is possible to obtain the Friedmann equations from the first law of thermodynamics and the continuity equation, using the following map relating the relative rate of expansion $H(t) = \dot{a}/a$, the temperature $T$ and the entropy $S$ of the apparent cosmological horizon \cite{Cai:2005ra,Danielsson:2004xw}:
\begin{equation}
H(t) = \left(\frac{\rho_{m} + \Lambda}{3}\right)^\frac{1}{2}\;, \qquad T = \frac{H(t)}{2\pi}\;,  \qquad S = \frac{\pi}{G H^2(t)}\;,
\end{equation}
where $\rho_{m}\propto t^{-2}$ is the matter density and $\Lambda$ the cosmological constant.
It is straightforward to realize that, as matter dilutes, the Hubble function $H(t)$ approaches the asymptotic value $H = \sqrt{\Lambda/3}$. Consequently, in appropriate units, the latter is the minimum temperature accessible to the system.

\begin{figure}[h!]
	\includegraphics[scale=0.38]{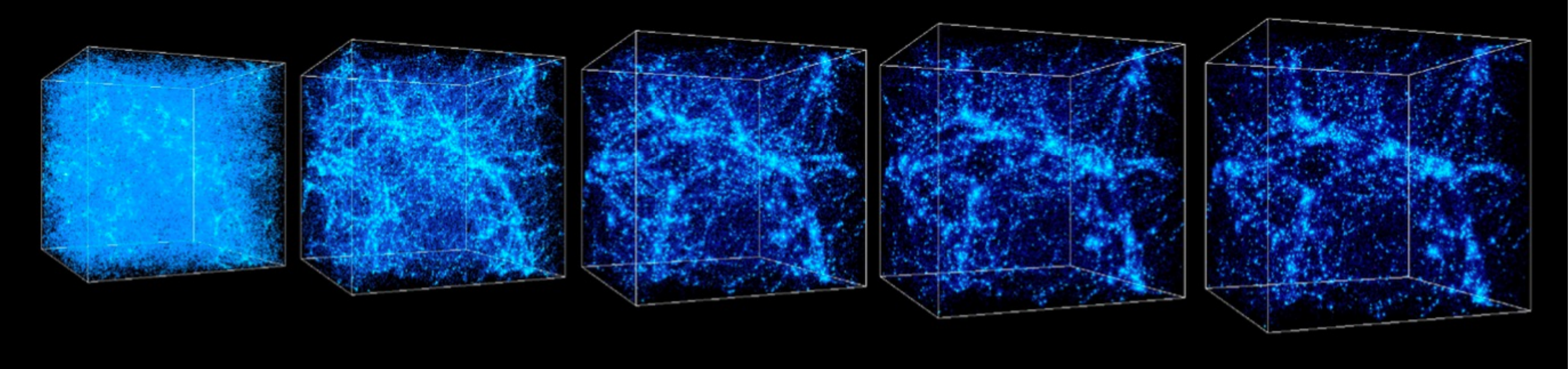}
	\caption{Simulated evolution of the Cosmic Web for a box of 43 million parsecs with $\Omega_m = 0.3$ and $\Omega_\Lambda = 0.7$, from left to right at redshift $z=9.8, z=2.97,\; z=0.99,\; z=0.5,\; z=0$. Simulations performed\footnote{Figure taken from \hyperlink{http://cosmicweb.uchicago.edu/filaments.html}{http://cosmicweb.uchicago.edu/filaments.html} } at the National Center for Supercomputer Applications by Andrey Kravtsov (The University of Chicago) and Anatoly Klypin (New Mexico State University).}
	\label{Cosmicweb}
\end{figure}

In a $\Lambda$CDM Universe, structures form in the matter dominated epoch, when the expansion rate decelerates enough for gravitational collapse to occur. This can be qualitatively understood in terms of the linear growth rate function \cite{Amendola:2015ksp}, defined as $D(a)/a$, where $D(a)$ is the growing mode of the density contrast $\delta = \delta\rho/\rho$. It is possible to express the growth rate using the following approximation \cite{Carroll}:
\begin{equation}\label{growthrate}
    \frac{D(a)}{a} \approx \frac{5}{2}\Omega_m\left[\Omega_m^{\frac{4}{7}} - \Omega_\Lambda +\left(1+\frac{\Omega_m}{2}\right)\left(1+\frac{\Omega_\Lambda}{70}\right)\right]^{-1} \; ,
\end{equation}
where $\Omega_m = \rho_m/3H^2$ and $\Omega_\Lambda = 1 - \Omega_m$. As long as $\Omega_\Lambda \ll 1$, the above equation gives $D(a)/a \approx 1$, which means that the gravitational collapse of linear perturbations is efficient despite the Universe's expansion. On the other hand, at late times, $\Omega_\Lambda$ grows significantly from 0 and the ratio $D(a)/a$ becomes smaller than unity. In other words, the repulsive force exercised by the cosmological constant $\Lambda$ will prevent the collapse to continue indefinitely, and eventually triggers the so called \textit{freeze out} of the growth of structures.  One can appreciate from Fig.\;\ref{Cosmicweb} that the distribution of structures does not change much from redshift $z \approx 0.5$ to $z=0$, which is precisely when we expect the contribution from Dark Energy to become significant (for reference, the densities of DE and DM become equal roughly at $z \approx 0.4$).
Thus, there is a temporal window in the evolution of the Universe during which the gravitational field due to the matter inhomogeneities reaches an almost stationary spatial configuration on sufficiently large scales (and it is a very reasonable question to ask whether it is a coincidence or not that this window is roughly today). 

Let us briefly summarize the main ideas we have presented so far:
\begin{itemize}
    \item Perturbations of the homogeneous and isotropic early Universe, if large enough, may undergo gravitational collapse described by the BIX geometry. Since the latter has the symmetry group $SU(2)$, the resulting object could possess an intrinsic, \textit{discrete} spin. We will assume, in the following, that the DM halos in the Universe originated in this way and have the above feature.
    \item There is a formal analogy between the Friedmann equations and the first law of thermodynamics for a system with temperature $T 
    \propto H(t)$ and entropy $S\propto H(t)^{-2}$.
    \item At late times, the distribution of structures on sufficiently large scales becomes almost stationary (more precisely, at rest with the Hubble flow). This occurs when the Cosmological Constant density dominates over the dark matter energy density, i.e.~when the Hubble factor becomes slowly varying and approaches a constant value.
\end{itemize}
Putting together these pieces, the following picture emerges: at relatively low redshift the Cosmic Web is equivalent to a (not really ordered) crystal in a thermal reservoir with temperature $T \propto H$. When $\Lambda$ dominates, the Hubble factor approaches a constant value and the temperature of the thermal bath is close to its minimum possible value. The crystal's components have intrinsic, discrete, magnetic spin, which is inherited from the BIX geometry underlying their process of structure formation. 
We argue that a suitable, simplified description of such a system is the Ising model \cite{Ising,Isingrev} with the following  Hamiltonian:
\begin{equation}\label{Isingham}
    \hat{\mathcal{H}} = -\sum_{<i,j>}^NJ_{ij}\hat{\sigma}_i\hat{\sigma}_j \:,
\end{equation}
where $J_{ij}$ is the interaction coupling, $N$ the number of fermions with spin $\sigma_i$. We will also make the following simplifying assumptions: the spins may take the values $\pm 1$ only ($\uparrow$ and $\downarrow$ respectively), the coupling $J_{ij} = J$ is constant and equal for all the spins, and in the sum $<i,j>$ in Eq.\eqref{Isingham} we can take into account only interactions between closest neighbours.
In the mean field approximation (which effectively describes the individual spins as non-interacting and sensitive only to the average spin of the system $\bar{\sigma}$), one obtains the following transcendental equation:
\begin{equation}\label{meanfieldIsing}
    \bar{\sigma} = Jn\tanh\left[\beta \bar{\sigma}\right]\,,
\end{equation}
where $\beta = \left(k_b T\right)^{-1}$ is the inverse temperature, $k_b$  the Boltzmann constant, and $n$  the number of closest neighbours.
When the above equation admits solutions different than $\bar{\sigma}=0$, a phase transition may occur and cause spontaneous magnetization.
Applied to the Cosmic Web, the above reasoning suggests that long-range correlations at redshift $z\leq 0.4$ may appear as long as Eq. \eqref{meanfieldIsing} admits non trivial solutions, and a second order phase transition take place. If this is the case, the Cosmic Web would behave like the gravitational analogue of a solid in a ferromagnetic state, with the ground state configuration having all the spins aligned in the same direction.

\begin{figure}[h!]
\includegraphics[scale=0.58]{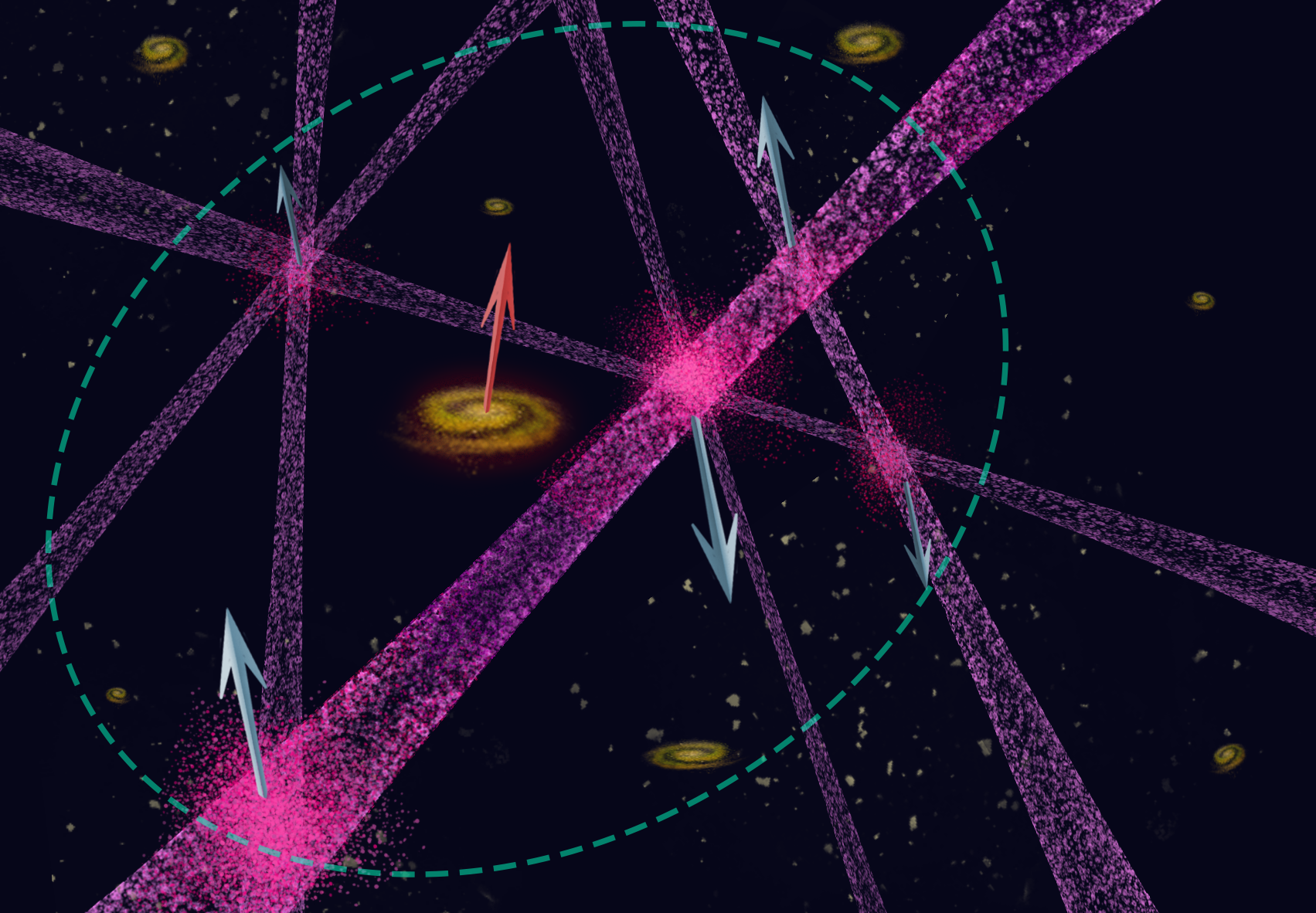}\;\;\;\includegraphics[scale=0.58]{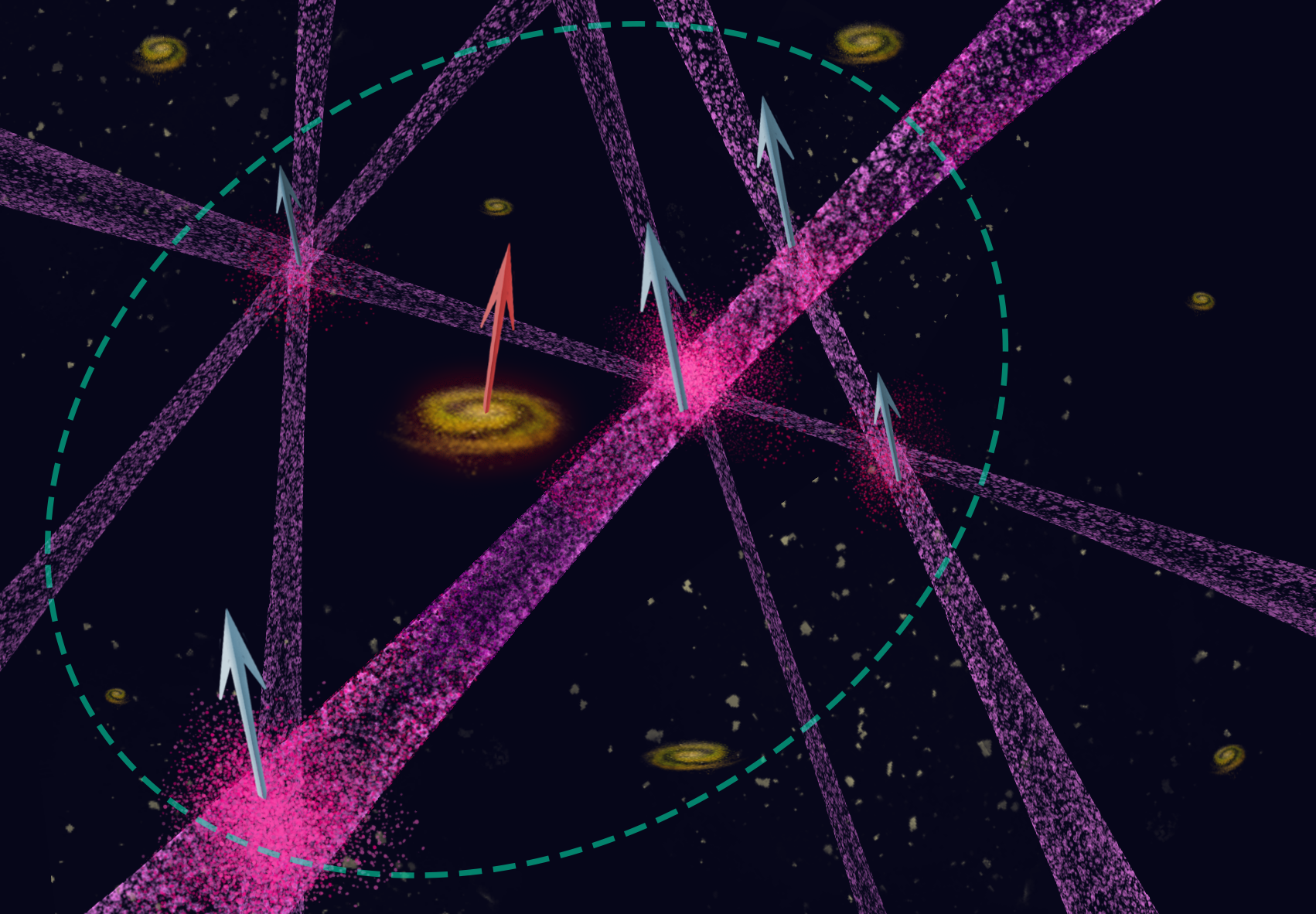}
	\caption{Figurative representation of a section of the Cosmic Web. The DM distribution is shown in purple, and baryonic matter in yellow. Notice that galaxies should be located within the DM halos, i.e.~within the purple stripes, but for illustrative purposes were depicted separately. The intrinsic discrete spin of the DM structures is represented with light blue arrows. The red arrow indicates the peculiar velocity of the galaxy at the center of the images. On the left, before the phase transition, the spins are randomly oriented and the net magnetization within the dashed blue curve vanishes. On the right, after the phase transition, the spins are uniformly oriented and $\bar{\sigma} \neq 0$. Accordingly, the galaxy in the center will feel the gravitational equivalent of the Lorentz force depending on the angle between the red and the light blue arrows.  
	Artwork designed by Yago Mongelli. }
	\label{jago}
\end{figure}

Before discussing the phenomenological implications arising from this picture, let us briefly quote Professor Weinberg for the last time:\footnote{\textit{The First Three Minutes} Publisher: Bantam Books, 1984, ISBN 10: 0553246828} \textit{Our mistake is not that we take our theories too seriously, but that we do not take them seriously enough. It is always hard to realize that these numbers and equations we play with at our desks have something to do with the real world.}
There is a remarkable analogy between the Maxwell equations and the gravitational ones in the weak field limit, usually referred to as \textit{Gravitoelectromagnetism} \cite{Clark:2000ff,thorne}. Following this analogy, it is possible to define the gravitational equivalent of the Lorentz force acting on a test particle of mass $m$:
\begin{equation}
\vec{F}=m\left[ \vec{E} + \vec{B}\times\vec{v}\right]\;,
\end{equation}
where $\vec{E}$ and $\vec{B}$ are the gravitoelectric and gravitomagnetic field, and $\vec{v}$ is the velocity of the test particle.
Due to the homogeneity and isotropy of the background, we expect these fields to average to zero on sufficiently large scales. On the other hand, if a phase transition occurs, the spins align and produce a non vanishing gravitomagnetic field $<\vec{B}>\; \propto \bar{\sigma}$. As a consequence, test particles with velocity of the same magnitude $|v|$ experience a force with intensity depending on their relative angle with respect to $\bar\sigma$, see Fig. \ref{jago}, and should therefore follow a dipole distribution.

Recently, there has been a lot of attention towards observational tests of the Cosmological Principle using a variety of different probes, see Refs. \cite{Rubart2013,Migkas:2021zdo,Bengaly:2017slg,Tiwari:2013vff,Singal:2014wra,Maartens:2017qoa} for some relevant examples. In general, these observations suggests that inconsistencies arise if we try to ascribe their dipole distribution solely to our peculiar motion with respect to the CMB rest frame. If these observations are not caused by unknown systematic, new physics at relatively low redshift should somehow be responsible either for significant bulk flows which extend beyond the scales predicted by $\Lambda$CDM simulations, or for a spatial variation of $H_0$ across the sky of order $\sim 1 - 10 \%$.\footnote{And therefore could potentially address the so called \textit{Hubble tension} \cite{Verde:2019ivm}.} In this perspective, a mechanism capable of producing long range correlations amongst structures in the late Universe is a valuable ally in defence of the Cosmological Principle.

\textit{Scarcely anyone who fully understand this theory can escape from its magic}, was the concluding remark used by Einstein in the preliminary announcement of his equations.\footnote{As quoted by S.Chandrasekhar in \cite{Chandra}, and reflected in W.Pauli's words in \cite{Pauli1958-PAUTOR}.} In retaliation, the laws of thermodynamics came to the world with the hope that time will eventually make sense of them: \textit{Who sees the future? Let us have free scope for all directions of research; away with dogmatism, either atomistic or anti-atomistic}\footnote{L. Boltzmann, from \textit{Lectures on Gas Theory}, translated by G.Brush \cite{Brush}.}. However, seems that time hasn't been kind: \textit{Thermodynamics is a funny subject. The first time you go through it, you don't understand it at all. The second time you go through it, you think you understand it, except for one or two points. The third time you go through it, you know you don't understand it, but by that time you are so used to the subject, it doesn't bother you anymore}.\footnote{Attributed to A. Sommerfeld.} Despite their differences there is, in our opinion, a higher level of elegance flowing from the yet mysterious overlap of General Relativity and Statistical Mechanics. Ultimately, this essay attempts to draw from this elegance in a yet unexplored but fascinating territory. It may have failed in doing so, but if this is the case we would like to remind Boltzmann's opinion in these regards: \textit{Elegance should be left to shoemakers and tailors}.\footnote{Attributed to L. Boltzmann, as reported by A. Einstein in Ref. \cite{ nla.cat-vn2298651}}

\clearpage
\section*{Acknowledgments}
We are deeply grateful to Cullan Howlett, Oliver F. Piattella, Pat Scott and Sunny Vagnozzi for useful discussions, comments and suggestions. We are also grateful to Yago Mongelli\footnote{yagomongelli@gmail.com} for his artistic contribution. 
LG and TMD acknowledge support from the Australian Government through the Australian Research Council Laureate Fellowship grant FL180100168.

\bibliographystyle{apsrev4-2.bst}
\bibliography{biblic.bib}

\end{document}